# Interface roughness transport in THz quantum cascade detectors

Emmanuel Lhuillier, Isabelle Ribet-Mohamed, Emmanuel Rosencher,
*ONERA, centre de Palaiseau, Chemin de la Hunière- FR 91761 Palaiseau cedex, France.*

Gilles Patriarche
*Laboratoire de Photonique et de Nanostructures, LPN/UPR20—CNRS Route de Nozay, 91460 Marcoussis, France.*

Amandine Buffaz, Vincent Berger
*Laboratoire Matériaux et Phénomènes Quantiques, Université Paris Diderot, CNRS UMR 7162, Bâtiment Condorcet, Case 7021, 75205 Paris cedex 13, France.*

Mathieu Carras
*Alcatel-Thales III-V Lab, Campus de Polytechnique, 1 Avenue A. Fresnel, 91761 Palaiseau cedex, France.*

PACS : 72.10Fk, 85.60Gz, 68.37Lp

Infrared Detectors based on a Quantum Cascade have been proposed to suppress the dark current which is identified as a limiting factor in Quantum Well Infrared Photodetectors. Those detectors have been mostly designed for the 3-5µm and 8-12µm range of wavelength. For detector operating in the THz range a complete change of regime of transport is expected since the photon energy is lower than the Longitudinal Optical (LO) phonon energy. Using a two dimensional code of transport we have identified Interface Roughness (IR) as the key interaction in such a structure. We have used scanning transmission electron microscopy (STEM) to evaluate the IR parameters ($\Delta$: magnitude of the roughness and $\xi$: mean distance between defects) instead of the classical mobility measurements. Finally, we used these parameters to study their influence on the resistance of the device.

In order to solve the issue of a too large dark current in Quantum Well Infrared Photodetectors (QWIPs), photovoltaic detectors inspired from Quantum Cascade Laser (QCL) and named Quantum Cascade Detectors (QCD) have been designed[1,2,3,4,5]. They consist in a given number of identical series of quantum wells. As in QWIPs, the infrared absorption occurs between two subbands separated by an energy corresponding to the targeted wavelength. However the excited electron is not extracted from the top subband by an external electric field but by resonant tunneling to the adjacent quantum well. It then cascades from subband to subband down to the next period's lower subband of the optical transition. Such detectors were initially designed for 3-5µm and 8-12µm range of wavelength. The involved energy differences between subbands and the operating temperatures associated to these spectral ranges are high enough for efficient longitudinal optical (LO) phonon emission and absorption. In previous studies[5], it has been identified as the parameter dominating the transport. However, this reasoning fails when THz energies and lower temperatures are concerned. In this particular case the main interaction has still to be identified.

In this letter, we first predict that the dominant interaction in THz QCDs is interface roughness (IR). We then use scanning transmission electron microscopy (STEM) measurements to extract the key parameters of IR: $\Delta$ the magnitude of the roughness and $\xi$ the mean distance between defects. In this study, we focus on a QCD specifically designed for the THz range. This device has been chosen not only because it was a home made design but also because it is the first real THz QCD as opposed to previous cascade detectors consisting in measuring a QCL at zero bias[2]. The sample studied is a multi period GaAs-Al$_{0.27}$Ga$_{0.73}$As wells and barriers. Each period is composed of five wells and barriers with the following width 52(b)/30(w)/48/50/44/60/50/68/54/. The first two wells of a period are doped with Si donor ($3\times 10^{17}$cm$^{-3}$) in the central third of the well. The photonic transitions $E_5$-$E_1$ and $E_4$-$E_1$ are expected to lead to a maximum of absorption at 70µm. This device, designed and grown by University Paris Diderot and the III-V Lab, will be presented and characterized in a future publication.

In the THz wavelength range, the dominant interaction is difficult to identify *a priori*. We used a hopping transport code[6] between 2D states based on the Fermi Golden Rule. Wave functions are evaluated using a two band kp model. All the most usual interactions met in GaAs are included: interaction between electron and LO phonon (LO), acoustical phonon (AC), alloy disorder (AL), interface roughness (IR), ionized impurities (II) and electrons (EE). We have evaluated the different relaxation rates in the cascade for the six processes. Calculations have been made for an electron hopping from the bottom of the initial subband ($K_i = 0$), for a temperature of 10K. The interaction parameters are given in ref 6. Table 1 shows the intracascade scattering rates, and between two different cascades. The numbering system of the levels is the same as in ref 5. The values show that interface roughness is the main interaction, the corresponding $1/\tau$ factor being at least of one order of magnitude above the others. As for LO phonon in the case of 3-12µm QCD, IR can be considered as the only interaction to evaluate the transport properties in THz QCDs. Now that these scattering rates have been evaluated, we can deduce the net scattering rates $G_{ij}$ between two levels. In particularly this calculation considers the population of each level:

$$G_{ij} = \int_0^\infty f_{FD}(E_i)(1-f_{FD}(E_f)) \cdot \frac{1}{\tau_{ij}(E_i-E_f)} dE_i$$

with $f_{FD}$ the Fermi Dirac distribution. We assume an equilibrium population for the electron, since the intrasubband relaxation is faster than the intersubband one.

| Process | $1/\tau_{54}$ | $1/\tau_{43}$ | $1/\tau_{32}$ | $1/\tau_{21}$ | $1/\tau_{53}$ | $1/\tau_{42}$ | $1/\tau_{31}$ | $1/\tau_{5'1}$ | $1/\tau_{4'1}$ |
|---|---|---|---|---|---|---|---|---|---|
| AL | $9.5\times10^{10}$ | $8.3\times10^9$ | $1\times10^{10}$ | $3.3\times10^9$ | $1.2\times10^{10}$ | $2.1\times10^8$ | $3.5\times10^8$ | $3\times10^{10}$ | $6.3\times10^{10}$ |
| **IR** | $\mathbf{4.2\times10^{12}}$ | $\mathbf{3.4\times10^{11}}$ | $\mathbf{4.5\times10^{11}}$ | $\mathbf{1.3\times10^{11}}$ | $\mathbf{4.1\times10^{11}}$ | $\mathbf{7.2\times10^9}$ | $\mathbf{1.4\times10^{10}}$ | $\mathbf{9.9\times10^{11}}$ | $\mathbf{2.4\times10^{12}}$ |
| II | $1.5\times10^{11}$ | $6.3\times10^9$ | $7.5\times10^{10}$ | $4.6\times10^{10}$ | $1.7\times10^9$ | $2.5\times10^8$ | $3.1\times10^9$ | $3.1\times10^7$ | $2.3\times10^8$ |
| EE | $1.4\times10^9$ | $2.4\times10^8$ | $5.3\times10^8$ | $5.1\times10^8$ | $2.4\times10^8$ | $1.3\times10^6$ | $4.6\times10^6$ | $1.3\times10^8$ | $8.3\times10^8$ |
| AC | $5.7\times10^9$ | $4.7\times10^8$ | $7.1\times10^8$ | $6\times10^7$ | $5.5\times10^8$ | $1.23\times10^7$ | $2.3\times10^7$ | $2.1\times10^9$ | $4.8\times10^9$ |
| LO | $3\times10^{-7}$ | $2.6\times10^{-8}$ | $4.3\times10^{-8}$ | $5.6\times10^{-8}$ | $2.8\times10^{-8}$ | $6.1\times10^{-10}$ | $1.3\times10^{-9}$ | $9.1\times10^{-8}$ | $2.1\times10^{-7}$ |

Tab. 1: Scattering rates (Hz) for different processes at $K_i=0$, under a low electric field (5749V.m$^{-1}$) and at T=10K. The numbering of the level allocates the number five to the highest level in energy and the number one to the lowest level of a cascade. The prime sign is used for the levels in the neighbouring cascade.

The $G_{ij}$ factors are then used for the calculation of $R_0A$ (where $R_0$ is the resistance of the pixel and $A$ the area of the pixel), via the relation $^5 R_0A = \dfrac{k_b T}{e^2 \sum_{i\in C}\sum_{j\in C'} G_{ij}}$, where C designates one cascade and C' the following one. We plot $R_0A$ for the different processes. We underline that because of the difficulty to evaluate separately the final energy for the electron-electron interaction and because this interaction is highly time consuming, we have chosen not to evaluate the $R_0A$ for this interaction.

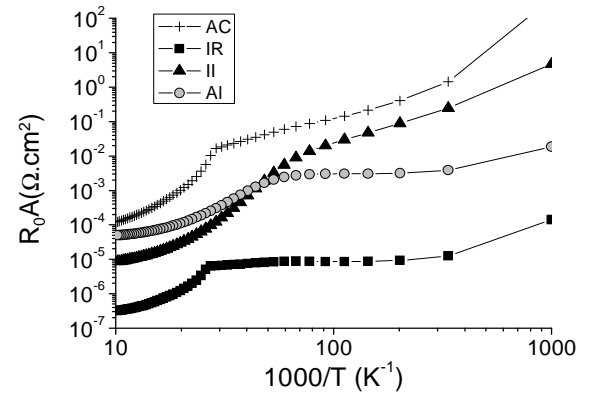

FIG. 1 $R_0A$ as a function of 1000/T for the different processes. The LO phonon process is not indicated on this graph since the associated $R_0A$ is much larger.

As predicted by the scattering rates, the limiting process is the IR. It leads to a low value of $R_0A$ ($10^{-5}$ $\Omega cm^2$, whereas $10^3$ $\Omega cm^2$ are often reported for 10 μm QCD). A way to improve the $R_0A$ value could consist in applying a higher electric field on the QCD, thus the transition i->i' would lose its elastic behaviour and the transport via interface roughness interaction would become less efficient. The increase of the $R_0A$ factor at high temperature (low 1000/T) may be attributed to the possibility of transition between levels with high energy in the cascade. For moderate temperature, the $R_0A$ presents a low dependence due to the fact that the initial Fermi factor can be considered as constant. At low temperatures, the increase of the $R_0A$ may be attributed to the population filling of the final subband.

In the following we only focus on the interface roughness interaction, since it prevails on the others at least from one order of magnitude. The evaluation of the net scattering rate allows us to conclude that: (i) The relaxation inside a cascade is done step by step. This is a major difference with the 10μm structure in which the LO phonon coupling allows the electron to jump more than one level. Here since the coupling is driven by an elastic process, the first neighbour jump prevails. (ii) Once an electron is promoted over the excited state 5, it tends to go back to the previous cascade ($G_{1'5}=8\times10^{16}$ $m^{-2}s^{-1}$ ~ $G_{5'4'}=3.6\times10^{17}$ $m^{-2}s^{-1}$). We can thus assume a drop of the Fermi level between two cascades.

In order to evaluate the parameter Δ (magnitude of the roughness) and ξ (mean distance between defects) most of the authors used parameters evaluated from mobility measurements[7,8]. We choose an alternative way by using STEM imaging, which gives direct access to Δ and ξ, contrary to mobility measurements in which those parameters are estimated from transport measurements. The sample is described in ref. 4 and corresponds to a 9μm QCD. As mentioned above, such a device is not driven by IR but the structure of this sample is quite similar to the one dedicated to the THz in terms of proportion of aluminium (34% vs 27%), which drives the IR. STEM images were acquired at 200 keV with a JEOL JEM 2200FS scanning transmission electron microscope equipped with a CEOS aberration-corrector.

HAADF-STEM images (also called high-resolution Z contrast images) were obtained with a half-angle probe of 30 mrad, the inner and outer half-angles of the annular detector (called upper-HAADF detector in this machine) were, respectively, 100 mrad and 170mrad, see FIG. 2. A cross section along the growth direction (FIG. 2 (b)) shows that the interfaces are not rigorously abrupt. We observed a gradient in the alloy concentration. The gradient of alloy concentration is as large as two or three monolayers (ML). A cross section along a well interface FIG. 2 (c)) gives information on the interface roughness. We observe an irregularity at the interface between GaAs and AlGaAs. This irregularity presents a ξ=10±3nm period (after averaging over several interfaces). The magnitude of the irregularity is of the same order of magnitude as the interface alloy gradient, meaning two or thee ML: Δ=0.6 to 0.9nm. Those values are similar to those found in the literature[6] ξ=6.5nm and Δ=0.3nm for typical GaAs heterostructure. Those measured parameters corroborate the dominant role play by IR a posteriori.

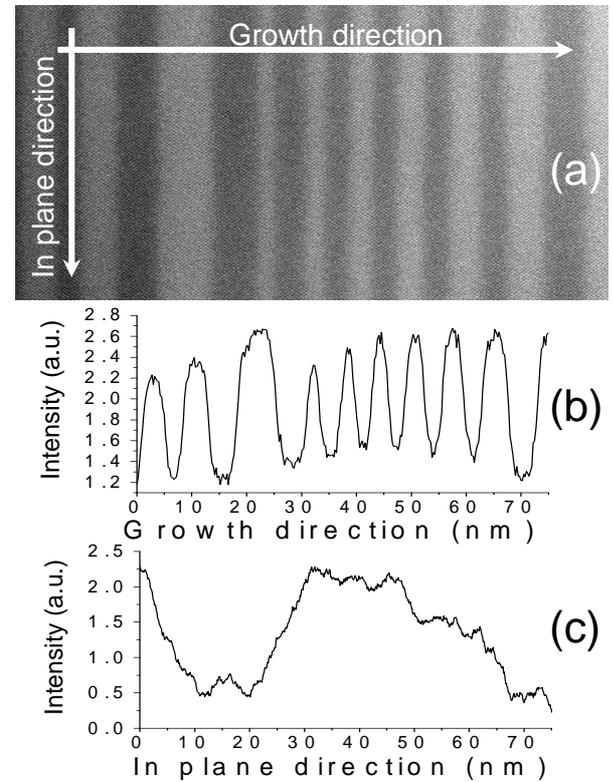

FIG. 2 (a) High resolution Z-contrast image of the QCD structure prepared by <110>. Corrected cross-section and its intensity profile along the growth direction (b) and along a well interface (c).

To finish we studied the influence of ξ and Δ on the $R_0A$ factor, for typical values around the experimental values. As expected, an increase of the magnitude of the IR leads to a lower $R_0A$. By increasing the roughness magnitude the $R_0A$ curve is shifted with a quadratic dependence toward Δ. The dependence of $R_0A$ upon the ξ parameter is not obvious since IR acts as a filter in momentum space for the electron, see FIG. 3. For the typical photon energy in our structure, an increase of ξ leads to a decrease of $R_0A$.

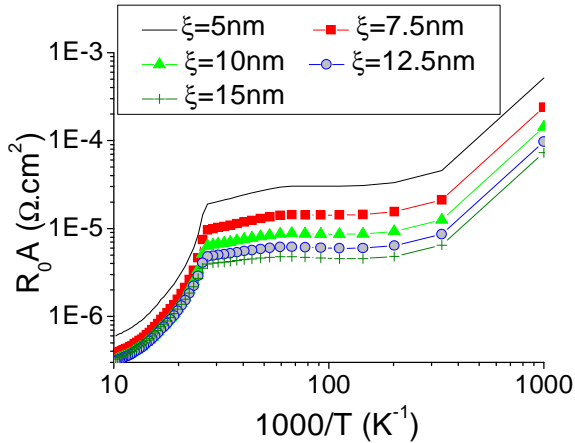

FIG. 3 $R_0A$ as a function of 1000/T, for Δ =2ML and different values of ξ.

To summarize we have identified that IR plays a key role in THz QCD. This interaction prevails over all the others at least from one order of magnitude. We used STEM imaging to estimate the IR characteristic parameters and we finally studied how those growth dependant parameters influence the performance of the QCD. A decrease of the number of levels into the cascade could reduce the number of interfaces and also limit the elastic behaviour of each transition, and thus improve the structure performances.

[1] V. Berger, French patent: Détecteurs à cascade quantique, 2001, National reference number 0109754.

[2] D. Hofstetter, M. Beck and J. Faist, Appl. Phys. Lett. **81**, 2683 (2002).

[3] L. Gendron, C. Koeniguer, V. Berger and X. Marcadet, Appl. Phys. Lett. **86**, 121116 (2005).

[4] L. Gendron, C. Koeniguer and V. Berger, Appl. Phys. Lett. **85**, 2824 (2004).

[5] C. Koeniguer, G. Dubois, A. Gomez and V. Berger, Phys. Rev. B **74**, 235325 (2006).

[6] E. Lhuillier, I. Ribet-Mohamed, A. Nedelcu, V. Berger and E. Rosencher, arXiv:0905.2062.

[7] G. Gottinger, A. Gold, G. Absteiter, G. Weimann and W. Schlapp, Europhysics letters **6**, 183 (1988).

[8] H. Sakaki, T. Noda, K. Hirakawa, M. Tanaka and T. Matsusue, Appl. Phys. Lett. **51**, 1934 (1987).